\documentclass[aps,prx,twocolumn,superscriptaddress,showpacs,floatfix,longbibliography]{revtex4-2}
\usepackage{amsmath,amssymb,wasysym,graphicx}
\pdfoutput=1
\usepackage[utf8]{inputenc}
\usepackage[T1]{fontenc}
\usepackage{xcolor}

\IfFileExists{newtxtext.sty}
   {\usepackage{newtxtext,newtxmath}}
   {\IfFileExists{stix.sty}
      {\usepackage{stix}}
      {\IfFileExists{mathptmx.sty}
      {\usepackage{mathptmx}}{} } }

\usepackage{textcomp}

\usepackage{bm}

\IfFileExists{siunitx.sty}{\usepackage{booktabs,siunitx}}{}

\usepackage{color}
\definecolor{LinkColor}{rgb}{0.256,0.439,0.588}
\usepackage{hyperref}
\hypersetup{
   colorlinks=true,
   citecolor=LinkColor,
   linkcolor=LinkColor,
   urlcolor=LinkColor
}

\usepackage{pifont}
\usepackage[normalem]{ulem}

\begin{document}

\title{Dirac fermions with plaquette interactions. II. $SU(4)$ phase diagram with Gross-Neveu criticality and quantum spin liquid}

\author{Yuan Da Liao}
%\email{ydliao@fudan.edu.cn}
\affiliation{State Key Laboratory of Surface Physics, Fudan University, Shanghai 200438, China}
\affiliation{Center for Field Theory and Particle Physics, Department of Physics, Fudan University, Shanghai 200433, China}

\author{Xiao Yan Xu}
\email{xiaoyanxu@sjtu.edu.cn}
\affiliation{Key Laboratory of Artificial Structures and Quantum Control (Ministry of Education), School of Physics and Astronomy, Shanghai Jiao Tong University, Shanghai 200240, China}

\author{Zi Yang Meng}
\email{zymeng@hku.cn}
\affiliation{Department of Physics and HKU-UCAS Joint Institute of Theoretical and Computational Physics, The University of Hong Kong, Pokfulam Road, Hong Kong SAR, China}

\author{Yang Qi}
\email{qiyang@fudan.edu.cn}
\affiliation{State Key Laboratory of Surface Physics, Fudan University, Shanghai 200438, China}
\affiliation{Center for Field Theory and Particle Physics, Department of Physics, Fudan University, Shanghai 200433, China}
\affiliation{Collaborative Innovation Center of Advanced Microstructures, Nanjing 210093, China}

\begin{abstract}
At sufficiently low temperatures, interacting electron systems tend to develop orders. Exceptions are quantum critical point (QCP) and quantum spin liquid (QSL), where fluctuations prevent the highly entangled quantum matter to an ordered state down to the lowest temperature. While the ramification of these states may have appeared in high-temperature superconductors, ultra-cold atoms, frustrated magnets and quantum moir\'e materials, their unbiased presence remain elusive in microscopic two-dimensional lattice models. Here, we show by means of large-scale quantum Monte Carlo simulations of correlated electrons on the $\pi$-flux square lattice subjected to plaquette Hubbard interaction, that a Gross-Neveu QCP separating massless Dirac fermions and a columnar valence bond solid at finite interaction, and a possible Dirac QSL at the infinite yet tractable interaction limit emerge in a coherent sequence. These unexpected novel quantum states reside in this simple-looking model, unifying ingredients including emergent symmetry, deconfined fractionalization and the dynamic coupling between emergent matter and gauge fields, will have profound implications both in quantum many-body theory and understanding of the aforementioned experimental systems.
\end{abstract}

\date{\today}

\maketitle

\section{Introduction}
The quantum mechanical description of the relativistic electron is attributed to Dirac, who revealed both its intrinsic angular momentum (the spin), with a half-integer quantum number $S = 1/2$ and the existence of its antiparticle, the positron~\cite{weinberg1995quantum}.
We now call these particle fermions and they obey the Fermi-Dirac statistics, which implies that two identical particles cannot occupy the same quantum mechanical state. In case of a vanishing rest mass, the energy of
Dirac fermions is a linear function of momentum.
Such massless Dirac fermions, when interacting with each other, give rise to interesting phenomena associated with quantum materials like graphene~\cite{novoselov2005two}, surfaces of topological insulators~\cite{chen2009experimental,zhang2009topological}, kagome metals~\cite{yeMassive2018,kangDirac2020} and twisted bilayer graphene~\cite{Bistritzer_TBG,cao2018correlated,cao2018unconventional,arora2020superconductivity,kerelsky2019maximized,andrei2020graphene,Yankowitz2019,Sharpe2019,lu2019superconductors,saito2020independent,stepanov2020untying,CaoYuan2020,polshyn2019large,xie2019spectroscopic,jiang2019charge,choi2019electronic,wong2020cascade,liu2020tunable,cao2020tunable,guinea2018electrostatic,liaoCorrelated2021,liaoValence2019,liaoCorrelation2021,xuKekule2018} and many other quantum moir\'e systems~\cite{saito2016superconductivity,wang2020correlated,Serlin2020,shen2020correlated,chen2019evidence,chen2019signatures,chen2020tunable,zhou2021superconductivity}.
The interplay between their Dirac relativistic dispersion and the onsite, extended and long-range Coulomb interactions these fermions experienced in the aforementioned materials is believed to be the magic ingredient that gives rise to the plethora of fascinating observed phenomena and has attracted broad attentions from communities encompassing quantum technology and devices and fundamental theories in condensed matter and high-energy physics.

At the theoretical front, model design, field theoretical analysis and large-scale numerical simulations have already provide valuable results on interacting $SU(2)$ Dirac fermions. Studies of Hubbard-like models on the honeycomb or $\pi$-flux square lattice suggested the emergence of exotic phases and phase transitions such as possible spin liquids~\cite{assaadPhase2005,mengQuantum2010,changQuantum2012,ouyangProjection2021},
valence bond solid (VBS) states~\cite{langDimerized2013,satoDirac2017,xuKekule2018,liaoValence2019}, quantum spin Hall states and superconductivity~\cite{hohenadlerQuantum2012,mengThe2014,wangPhases2021,wangDoping2021,liuMetallic2022} and Gross-Neveu and deconfined quantum critical points (QCP)~\cite{otsukaUniversal2016,toldinFermionic2015,langQuantum2019,liaoValence2019,liuSuperconductivity2019,liDeconfined2019,liuDesigner2020,moitabaChiral2021,janssenConfinement2020,wangDoping2021,schwabNematic2022,liaoGross2022,zhuQuantum2022} at or near half-filling per site per degree of freedom.
It's also worth to note that one always find the VBS win over the AFM in such models when $N\ge4$~\cite{zhouMott2016}.

It is anticipated that, by assigning more degrees of freedom to the Dirac fermions and with the extended interaction beyond the onsite Hubbard type, the system will acquire larger parameter space and exhibit more interesting behavior. Experimentally, the fermionic alkaline cold-atom arrays could realize the $SU(N)$ group with $N$ upto 10 and substantial progress has been made with magnetism and Mott transition realized~\cite{gorshkovTwo2010,miguelUltracold2014}. Signatures of quantum spin liquid and topologically ordered state of matter have been reported in Rydberg atom arrays where the long-range interaction is dominate~\cite{Semeghini21,Roushan21,Samajdar:2020hsw,yanTriangular2022}. Electrons in the twisted bilayer graphene and other quantum moir\'e materials are naturally bestowed with more degrees of freedom such as layer, valley and band and subject to extended and even truly long-range Coulomb interactions~\cite{Bistritzer_TBG,cao2018correlated,cao2018unconventional,arora2020superconductivity,kerelsky2019maximized,andrei2020graphene,Yankowitz2019,Sharpe2019,lu2019superconductors,zhangMomentum2021,panDynamical2022,ouyangProjection2021,zhangSuperconductivity2021,panSport2022}. However, the lack of appropriate model design and the unbiased methodologies prohibits the comprehensive understanding of interacting $SU(N)$ Dirac fermions subjected to extended interactions compared with their $SU(2)$ cousins.

\begin{figure*}[htp!]
\includegraphics[width=2\columnwidth]{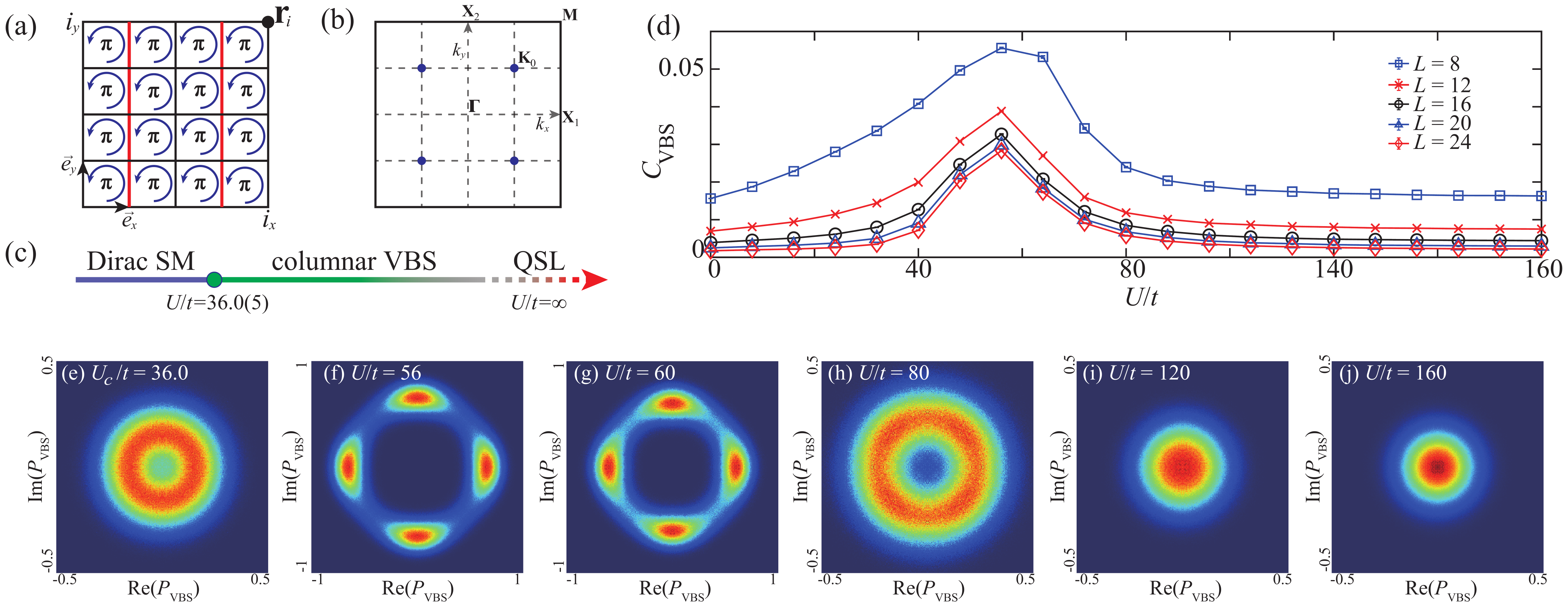}
\caption{(a) Dirac electrons on the square lattice with $\pi$-flux hopping. Black and red solid lines denote the hopping amplitude $t$ and $-t$.  $\mathbf{r}_i \equiv i_{x}\vec{e}_{x}+i_{y}\vec{e}_{y}$ is the position of site $i$, where $\vec{e}_x$ and $\vec{e}_y$ are the unit vector along x and y direction, respectively. (b) The square is the first BZ of the square lattice. The blue solid points represent the positions of Dirac cones at $\mathbf{K}_0 = \left( \pm\frac{\pi}{2},\pm\frac{\pi}{2}\right)$. High symmetry points $\mathbf{\Gamma}=(0,0)$, $\mathbf{X}_1=(\pi,0)$, $\mathbf{X}_2=(0,\pi)$ and $\mathbf{M}=(\pi,\pi)$ are denoted. (c) The ground state phase diagram of Eq.~\eqref{eq:model} obtained from PQMC and IQMC. As explained in the text, when $U<U_c/t=36.0(5)$ the system is inside a Dirac semimetal phase, and the Gross-Neveu chiral XY QCP is at $U_c$, and when $U>U_c$ there is a gradual change from cVBS to QSL as $U$ approching the $\infty$ limit. (d) Structure factor of cVBS obtained from PQMC as a function of interaction strength for different linear system sizes $L$. (e)-(j) the cVBS order parameter histogram $\arg(P_{\mathrm{VBS}})$ obtained from PQMC for $L=24$ system at $U/t=36.0, 56, 60, 80, 120$ and 160, respectively. At the Gross-Neveu QCP (e), there is an emergent U(1) symmetry. Well inside the cVBS phase (f) and (g), the $\arg(P_{\mathrm{VBS}})=(0, \frac{\pi}{2}, \pi, \frac{3 \pi}{2})$ due to the four-folded rotational symmetry breaking of the state. But as $U$ approaches the $\infty$ limit, (h), (i) and (j), the U(1) symmetry is restored at $U/t=80$ and the order parameter vanishes on this finite size at finite $U/t=120$ and 160.}
	\label{fig:phasediagram}
\end{figure*}

This work will address such lack of concrete knowledge in a decisive manner. Here we show, by means of three different yet complementary quantum Monte Carlo (QMC) simulation techniques on correlated $SU(4)$ Dirac fermions on the $\pi$-flux square lattice with extended plaquette interaction: a Gross-Neveu QCP separating massless Dirac fermions and a columnar VBS (cVBS) at finite interaction strength, and a Dirac quantum spin liquid (QSL)~\cite{ranProjected2007,xuMonte2019} at the infinite interaction limit emerge in a coherent sequence. We find such unexpected sequence of novel quantum states in the simple-looking model, unify the key ingredients including emergent continuous symmetry~\cite{xuKekule2018,liaoValence2019,janssenConfinement2020,liaoGross2022}, deconfined fractionalization~\cite{xuMonte2019,maDynamical2018,maRole2019} and the dynamic coupling between emergent matter and gauge fields~\cite{heQuantum2016,xuMonte2019,wangDynamics2019} and therefore provide a solid foundation for the future exploration of the novel quantum matter originated from the interplay of the low-energy relativistic dispersion and strong extended and long-range interactions, resonating with the aforementioned experiments and quantum materials and the original insight from Dirac.

\section{ Model and method}
We investigate a $SU(4)$ plaquette Hubbard model at half-filling on the $\pi$-flux  square lattice with the Hamiltonian
\begin{equation}
\label{eq:model}
	H=-\sum_{\langle i j\rangle, \alpha} t_{i j}\left(c_{i \alpha}^{\dagger} c_{j \alpha}+\text { H.c. }\right)+U \sum_{\square}\left(n_{\square}-2\right)^{2},
\end{equation}
where $\langle i j\rangle$ represent the nearest neighbors,
$c_{i \alpha}^{\dagger}$ and $c_{i \alpha}$ are creation and annihilation operators for fermions on site $i$ with flavor indices $\alpha\in [1,4]$ with $SU(4)$ symmetry,
$n_{\square}  \equiv \frac{1}{4} \sum_{i \in \square} n_{i} $ denotes the extended particle number operator at $\square$-plaquette, with $n_{i}=\sum_{\alpha=1}^{4} c_{i \alpha}^{\dagger} c_{i \alpha}$
and $\langle n_{\square} \rangle =2$ at half-filling,
$U$ is the tunable plaquette repulsive interaction strength.

As shown in Fig.~\ref{fig:phasediagram}~(a), we set the $\pi$-flux hopping amplitude $t_{i,i+\vec{e}_{x}} = t$  and  $t_{i,i+\vec{e}_{y}} = (-1)^{i_x} t$ respectively, where the position of site $i$ is defined as $\mathbf{r}_i = i_x\vec{e}_{x}+i_y\vec{e}_{y}$.  As discussed in Ref.~\cite{liaoGross2022}, the positions of Dirac cones and the way of folding of the first Brillouin zone (BZ) will change with different gauge choice of $t_{ij}$, however, the distance between two Dirac cones does not.
Thus we could perform analysis in the original square lattice BZ, as shown in Fig.~\ref{fig:phasediagram}~(b). We set $t=1$ as the energy unit throughout the article and the
$\pi$-flux hopping term on square lattice gives rise to the dispersion
$\epsilon(\mathbf{k}) = \pm 2 t \sqrt{\cos^2 k_x + \cos^2 k_y} $, with
the Dirac cones located at momentum $\mathbf{K}_0=\left(\pm \frac{\pi}{2}, \pm\frac{\pi}{2}\right)$, which results in the Dirac semimetal state in the weak coupling region. As the interaction strength increases, it is expected that the Dirac cones will be gapped out (the relativistic Dirac fermions will acquire interaction-generated mass) and an insulating phase stemmed from Mott physics will have the upper hand~\cite{mottBasis1949,sorellaSemi1992,caiQuantum2013,wangCompeting2014}.
In our model, the plaquette interaction term $U$ naturally contains the onsite, first and second nearest neighbor repulsions.
Similar to the previous studies~\cite{xuKekule2018,liaoValence2019,liaoCorrelation2021,liaoCorrelated2021,liaoGross2022}, the Mott insulator phase will require a relatively larger $U$ to occur, and in the infinite $U$ limit, a QSL phase may emerge~\cite{ouyangProjection2021}.

With the help of particle-hole symmetry, our QMC simulation has no sign problem~\cite{wuSufficient2005,caiPomeranchuk2013,panSign2022}.
To obtain the truly unbiased numerical results, we employ three different but complementary QMC algorithms to yield a consistent picture.
Most of data are obtained by projector QMC (PQMC)~\cite{assaadWorld-line2008}, which is suitable to investigate the ground state properties.
In addition, we also apply the finite temperature QMC (FTQMC) to investigate the thermodynamic and dynamic properties of different phases.
What's more, a new approach of QMC developed by one of us~\cite{ouyangProjection2021} that can perform simulations at the $U=\infty$ for zero and finite temperature is used here to study the nature of the model in Eq.~\eqref{eq:model} in the strong coupling limit, and we would like to call this method IQMC for short.
The details of these algorithms can be found in previous studies~\cite{xuKekule2018,liaoValence2019,liaoCorrelation2021,liaoCorrelated2021,liaoGross2022,ouyangProjection2021} and Sec.~\ref{app:A} in the Appendix.
We just mention to set projection time $\Theta t = L$ for equal-time measurement, $\Theta t = L+20$ for imaginary-time measurement and discrete the time slice $\Delta\tau=0.1$ in PQMC and IQMC method, and $\Delta\tau=0.05$ for FTQMC method, and we have simulated the square lattice system with $N=L^2$ sites and the linear size $L=8,12,16,20,24,28$.

\section{ Phase diagram }
The ground state phase diagram of our model obtained from QMC is shown in Fig.~\ref{fig:phasediagram}~(c).
In the weak coupling region, the model features a Dirac semimetal (SM) state due to the stability of relativistic Dirac fermions.
Increasing $U$, the Dirac SM transits into an Mott insulator state with cVBS order via a Gross-Neveu QCP at $U_c/t=36.0(5)$.
Surprisingly, we find as further approaching the strong coupling limit at $U=\infty$, the cVBS gradually evolves into a possible Dirac QSL. From our thermodynamic and dynamic measurements, the Dirac QSL is consistent with the state of emergent spinons with (again) massless Dirac relativistic dispersion coupled with dynamic [possibly U(1)] gauge field. Such a novel state of matter is at the heart of many intriguing quantum many-body phenomena: in condensed matter, the $(2+1)$D field theories with a compact U(1) gauge field coupled to relativistic Dirac fermions often serve as the low-energy effective field theories for high-temperature superconductors~\cite{leeDoping2006,leeU12005}, algebraic spin liquid~\cite{hermeleAlgebraic2005,kimMassless1997,ranProjected2007,hermeleErratum2007,xuMonte2019,wenTheroy1996,dupuisAnomalous2021,calveraTheory2021} and the deconfined quantum criticality~\cite{maDynamical2018,senthilQuantum2004,qinDuality2017}; in high-energy physics, the mechanism of quark confinement in gauge theories with dynamical fermions such as quantum chromodynamics (QCD) is among the most difficult subjects, and the absence or presence of a deconfined phase in 3D compact quantum electrodynamics (cQED3) coupled to massless Dirac fermions has attracted a
lot of attention and remains unsolved to this day~\cite{fiebigMonopoles1990,herbutPermanent2003,hermeleAlgebraic2005,nogueiraCompact2008,karthikNumerical2019,karthikQED32020,calveraTheory2021,albayrakBootstrapping2022}.

Following our previous study of $SU(2)$ $\pi$-flux square lattice extended Hubbard model~\cite{liaoGross2022}, to confirm the VBS order, we define the VBS structure factor
\begin{equation}
C_\text{VBS}^{\mathbf{e}} (\mathbf{k}, L) = \frac{1}{L^4} \sum_{i,j} e^{i \mathbf{k} \cdot\left(\mathbf{r}_{i}-\mathbf{r}_{j}\right)} \left\langle B^{\mathbf{e}}_{i} B^{\mathbf{e}}_{j}\right\rangle,
\end{equation}
where
$ B^{\mathbf{e}}_{i} = \frac{1}{4} \sum_{\alpha=1}^{4}\left( t_{i, i+\mathbf{e}}  c_{i, \alpha}^{\dagger} c_{i+\mathbf{e}, \alpha}+ \text {H.c.} \right)$ are gauge invariant bond operators with $\mathbf{e}$ standing for $\vec{e}_x$ or $\vec{e}_y$.
For cVBS order, $C_\text{VBS}^{\vec{e}_x}(\mathbf{k}, L)$ is peaked at momentum $\mathbf{X}_1=(\pi,0)$ and $C_\text{VBS}^{\vec{e}_y}(\mathbf{k}, L)$ is peaked at momentum $\mathbf{X}_2=(0,\pi)$.
Since the perfect cVBS order has $Z_4$ degeneracy on square lattice, $C_\text{VBS}^{\vec{e}_x}(\mathbf{X}_1, L)$ should be equivalent to $C_\text{VBS}^{\vec{e}_y}(\mathbf{X}_2, L)$ in ideal QMC simulations.
Consequently, we can define the square of the  cVBS order parameter as
\begin{equation}
 C_\text{VBS}(L)= C_\text{VBS}^{\vec{e}_x}(\mathbf{X}_1, L)+C_\text{VBS}^{\vec{e}_y}(\mathbf{X}_2, L).
 \label{eq:eq3}
\end{equation}
As shown in Fig.~\ref{fig:phasediagram}~(d), as tuning $U/t$ from $0$ to $160$, $C_\text{VBS}$ first increases then decreases, which means our model transits from  Dirac SM to cVBS order, then cVBS order becomes weaker and gradually evolves into a possible Dirac QSL phase at the $U=\infty$ limit.
We also calculate the square of antiferromagnetic (AF) order parameter and extrapolate it to the thermodynamic limit (shown in Sec.~\ref{app:B} of Appendix) and do not find AF order in all parameter region simulated.
These results are summarized as the phase diagram in Fig.~\ref{fig:phasediagram} (c), and we now discuss in detail the sequence of phases and phase transitions along the $U/t$ parameter path.

\begin{figure}[htp!]
	\includegraphics[width=\columnwidth]{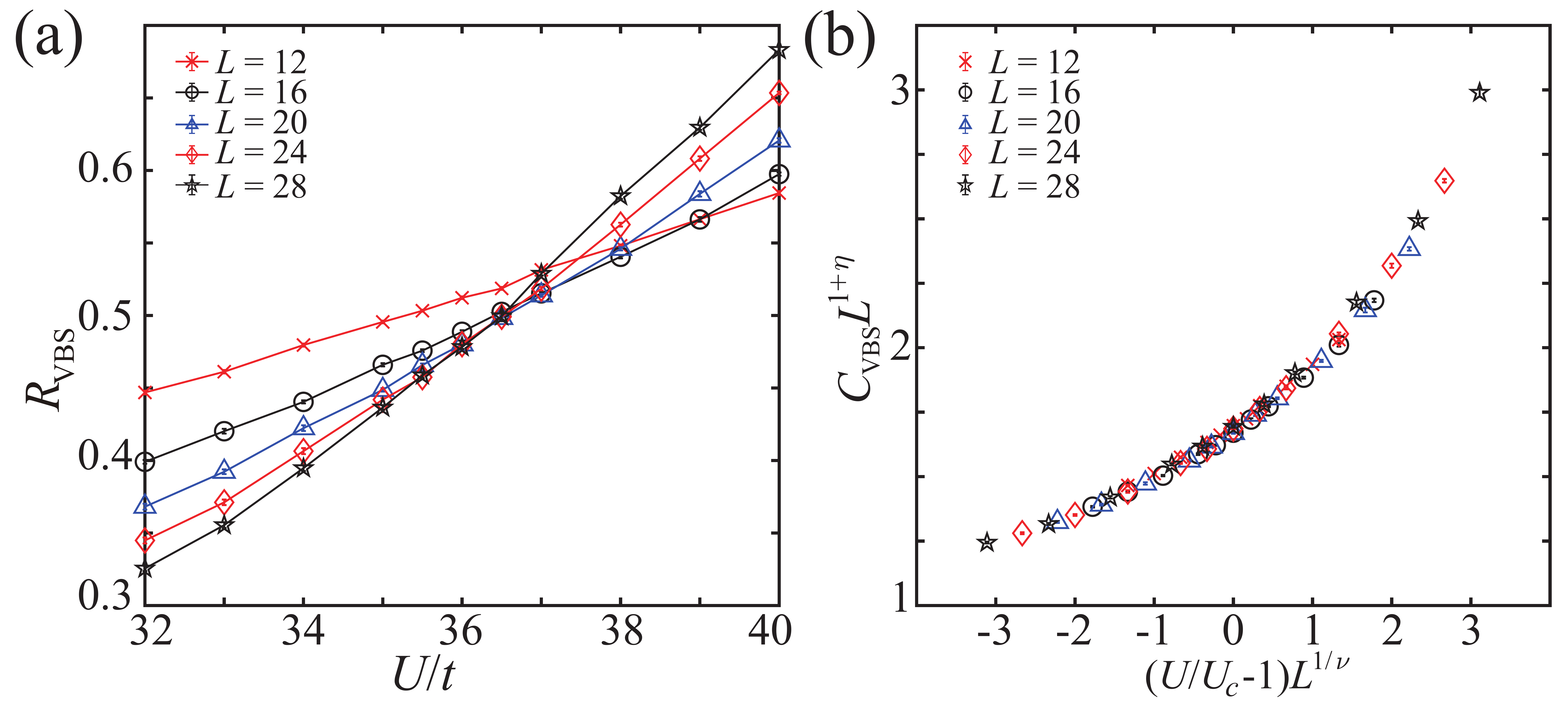}
	\caption{(a) Correlation ratio of VBS order in Eq.~\eqref{eq:eq4}, the crossing point is the Dirac SM to cVBS Gross-Neveu QCP. (b) Data collapse of VBS structure factor. The data cross and collapse give an estimate of the position of QCP at $U_c/t=36.0(5)$ with the critical exponents $\nu=1.00(5)$ and $\eta=0.86(4)$. (c) The histogram of $P_\text{VBS}$ at $U_c$ for $L=24$. There is an emergent $U(1)$ symmetry associated with the Gross-Neveu transition. (d) The histogram of $P_\text{VBS}$ at $U>U_c$ for $L=24$. The angular distribution is consistent with cVBS order. 
	These data are obtained from PQMC method.}
	\label{fig:SM-VBS}
\end{figure}

\section{Gross-Neveu transition and the Dirac QSL}
We first focus on the Gross-Neveu QCP between the Dirac SM and cVBS state. To locate the corresponding QCP, we define the correlation ratio of cVBS order as
\begin{equation}
R_\text{VBS}(L)=1-\frac{C_\text{VBS}^{\vec{e}_x}(\mathbf{X}_1+d \mathbf{q}, L)}{2C_\text{VBS}^{\vec{e}_x}(\mathbf{X}_1, L)}-\frac{C_\text{VBS}^{\vec{e}_y}(\mathbf{X}_2+d \mathbf{q}, L)}{2C_\text{VBS}^{\vec{e}_y}(\mathbf{X}_2, L)},
\label{eq:eq4}
\end{equation}
where $d\mathbf{q}$ is the smallest momentum in finite-size BZ.
$R_\text{VBS}(L)$ will approach to $1$ (0) in an ordered (disordered) phase. This dimensionless quantity is scale invariant at the QCP for sufficiently large system size~\cite{langDimerized2013,pujariInteraction2016,satoDirac2017,langQuantum2019,liaoGross2022}, which renders a crossing point  $U_c/t=36.0(5)$ as the position of the QCP among different $L$, shown in Fig.~\ref{fig:SM-VBS} (a).
Near the QCP, the cVBS structure factor should obey the scaling relation $C_\text{VBS}(L) = L^{-z-\eta} f \left(L^{1 / v}\left(U-U_{c}\right) / U_{c}\right)$, where $f$ is the scaling function, $z$ is the dynamical exponent and should be set as $1$ for relativistic Dirac fermions.
With this scaling relation, we collapse $C_\text{VBS}(L)$, as shown in Fig.~\ref{fig:SM-VBS}(b), and extract the universal critical exponents $\nu=1.00(5)$ and $\eta=0.86(4)$.
In principle, there are two kinds of VBS order, cVBS and plaquette VBS (pVBS), that share the same order parameter.
To further verify the cVBS order is in our model, we define the order parameter histogram
\begin{equation}
\begin{aligned}
&P_{1}=\left(1 / L^{2}\right) \sum_{i}\left(B_{i}^{\vec{e}_{y}}+\omega B_{i}^{-\vec{e}_{x}}+\omega^{2} B_{i}^{-\vec{e}_{y}}+\omega^{3} B_{i}^{\vec{e}_{x}}\right) e^{i \mathbf{X}_{1} \cdot \mathbf{r}_{i}}, \\
&P_{2}=\left(1 / L^{2}\right) \sum_{i}\left(B_{i}^{\vec{e}_{y}}+\omega B_{i}^{-\vec{e}_{x}}+\omega^{2} B_{i}^{-\vec{e}_{y}}+\omega^{3} B_{i}^{\vec{e}_{x}}\right) e^{i \mathbf{X}_{2} \cdot \mathbf{r}_{i}}, \\
&P_{\mathrm{VBS}}=P_{1}+P_{2}
\end{aligned}
\end{equation}
with $\omega=i$. For cVBS, the angular distribution of $P_{\mathrm{VBS}}$ will peak at the angles $\arg(P_{\mathrm{VBS}})=(0, \frac{\pi}{2}, \pi, \frac{3 \pi}{2})$,
while for pVBS, it will peak at $\arg(P_{\mathrm{VBS}})=(\frac{\pi}{4}, \frac{3 \pi}{4}, \frac{5 \pi}{4}, \frac{7 \pi}{4})$.
Our numerical results, as shown in Fig.~\ref{fig:phasediagram}(f) and (g) for $U/t=56,60$, confirm the cVBS.
More interestingly, as shown in Fig.~\ref{fig:phasediagram}(e) for $U/t=36$, there is an emergent $U(1)$ symmetry at $U_c$, which suggests that the corresponding QCP should be described by the 3D $N=4$ chiral Gross-Neveu XY universality class~\cite{schererGauge2016,liFermion-induced2017,classenFluctuation2017,torresFermion2018,janssenConfinement2020,wangDynamics2019}. Our extracted critical exponent $\nu$ and $\eta$ are comparable with the previous QMC~\cite{zhouMott2018,liFermion-induced2017,liaoValence2019} and $4-\epsilon$ expansion~\cite{rosensteinCritical1993,zerfFour2017} results on the same universality class.

\begin{figure}[htp!]
	\includegraphics[width=\columnwidth]{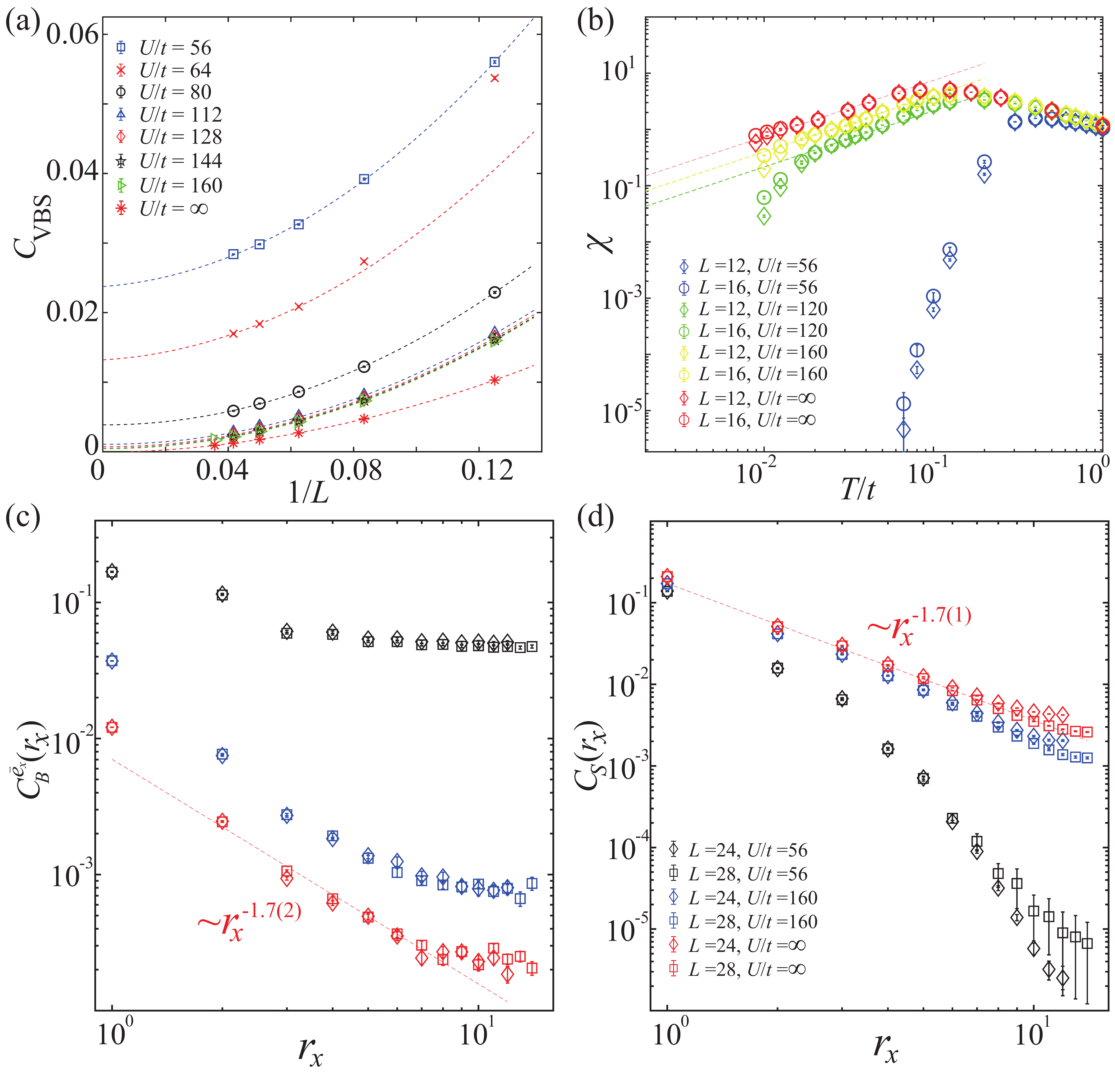}
	\caption{(a) $1/L$ extrapolation of $C_\text{VBS}$ obtained from PQMC and IQMC. The dash lines are quadratic polynomial fittings through the data points. As $U\to\infty$ the cVBS order extracts to vanishing small values. (b) Log-log plot of spin susceptibility $\chi$ as function of $T$ for different $L$ and $U$ obtained from FTQMC and IQMC. The dash lines, drawn as guide to the eyes, represent the linear functions $\chi=aT$ for different constant $a$.
		(c) Log-log plots of $C^{\vec{e}_{x}}_B (r_x)$ and (d) $C_{S}(r_x)$  as function of $r_x$ for different $L$ and $U$ obtained from PQMC. They have the same power-law decay as $1/r_x^{1.7}$ at $U=\infty$. The linear fitting with least squares method are used.}
	\label{fig:largeU}
\end{figure}

Next, we move on to the stronger interaction regime.
As shown in Fig.~\ref{fig:largeU}~(a), we extrapolate the cVBS structure factors to the thermodynamic limit for different $U>U_c$.
We notice that the VBS order becomes weaker as increasing $U$ and almost vanishes with the system sizes accessed for finite $U$ QMC around $U/t=160$. Our $U=\infty$ QMC simulations consistently reveal the $C_\text{VBS}$ extrapolate to $0$ at the strong interaction limit (the absence of the AF order on entire phase diagram is shown in Sec.~\ref{app:B} of the Appendix), which point to an emergent QSL state.
According to the theory of algebraic QSL with Dirac spinons coupled to U(1) gauge field~\cite{hermeleAlgebraic2005,hermeleErratum2007}, besides the absence of AF and VBS order parameters, the bond operator correlation
\begin{equation}
C^{\vec{e}_{x}}_B (\mathbf{r}) \equiv e^{-i \mathbf{X}_1 \cdot \mathbf{r} }\left(\left\langle B^{\vec{e}_{x}}_i B^{\vec{e}_{x}}_j\right\rangle-\left\langle B^{\vec{e}_{x}}_i\right\rangle\left\langle B^{\vec{e}_{x}}_j\right\rangle\right)
\end{equation}
and the AF (staggered) spin correlation
\begin{equation}
C_{S}(\mathbf{r}) \equiv e^{-i\mathbf{M}\cdot\mathbf{r}}\sum_{\alpha,\beta}\langle S^{\alpha}_{\beta}(i) S^{\beta}_{\alpha}(j)\rangle
\end{equation}
should decay algebraically at large separations.
In the above equations, $\mathbf{r}=\mathbf{r}_i -\mathbf{r}_j$ is the relative distance, $S^{\alpha}_{\beta}(i)=c^{\dagger}_{i,\alpha}c_{i,\beta}-\frac{\delta_{\alpha,\beta}}{4}\sum_{\gamma}c^{\dagger}_{i,\gamma}c_{i,\gamma}$ are the $SU(4)$ spin full operators with $\alpha,\beta,\gamma \in [1,4]$.
For both finite and $U=\infty$, $C^{\vec{e}_{x}}_B (\mathbf{r})$ and $C_{S}(\mathbf{r})$ with largest system sizes are shown in Fig.~\ref{fig:largeU}~(c) and (d), and we indeed numerically observe the algebraical behaviors in the correlation functions as $U\to\infty$. These are strong evidence of the robust existence of QSL. Most interestingly, we find the two correlation functions acquire the same power-law decay within errorbar, i.e. $\sim 1/r^{1.7(2)}$. This is a strong indication that, the QSL phase shall be understood with a low-lying effective theory with emergent relativistic Dirac spinons coupled with dynamic U(1) gauge field, as only in this way, the kinetic bond and $SU(4)$ spin correlation functions, although have different scaling dimension at the bare operator level, actually describe the correlation of the same effective degrees of freedom in low-energy effective field theory multiplet~\cite{hermeleAlgebraic2005,hermeleErratum2007,maDynamical2018,maRole2019,nahumDeconfined2015,nahumEmergent2015}.

Furthermore, the decay power indicates the QSL state is unlikely to be a $\mathbb Z_2$-Dirac QSL~\cite{AssaadZ2DSL2016,GazitZ2DSL2017}, because the $\mathbb Z_2$ gauge field does not have gapless excitations, and thus does not modify the decay power of free Dirac fermions, which would be 4.
However, the decay power is much smaller than previous DQMC and large-$N$ studies~\cite{hermeleAlgebraic2005,hermeleErratum2007,xuMonte2019} of U(1) Dirac QSL (which are both larger than 3).
Possible explainations of this discrepancy include finite-size effects, a new QSL state being realized, or that the $U\rightarrow\infty$ limit is a critical point between the VBS phase and a QSL phase instead of the QSL phase itself.
We leave this to future works.
Last, we notice that a recent work~\cite{calveraTheory2021} shows the monopole operator may be relevant in U(1) Dirac-QSLs and may lead to instabilities.
This may be related to the fact that we only observe QSL in the $U\rightarrow\infty$ limit: it is possible that an effective local constraint enforced by the infinite interaction helps stablize the QSL phase.
We leave detail studies on this issue to future works.

The cVBS order parameter histograms $\arg(P_{\mathrm{VBS}})$, on the other hand, clearly demonstrate the vanishing of the cVBS order as a function of $U$. As shown in Fig.~\ref{fig:phasediagram} (h), (i) and (j), the evolution of $\arg(P_{\mathrm{VBS}})$ for $L=24$ and $U/t=80, 120$ and 160, it is clear that the cVBS order parameter becomes weaker and loses the $Z_4$ anisotropy already at $U/t=80$ and completely vanishes on this finite size at $U/t=120$ and 160, consistent with the phase diagram in Fig.~\ref{fig:phasediagram} (c).

\begin{figure*}[htp!]
	\includegraphics[width=\textwidth]{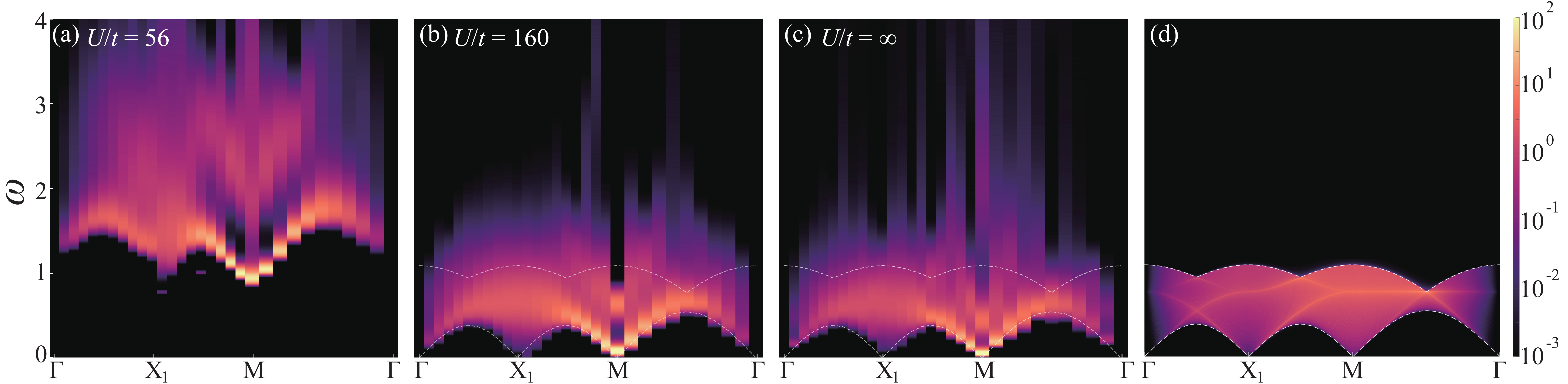}
	\caption{Dynamic spin spectra obtained from QMC-SAC calculations as a sequence inside the cVBS phase, towards and inside the Dirac QSL with $L = 20$. (a) is inside the cVBS phase with $U/t=56$, obtained from PQMC, (b) is close to the $U=\infty$ limit with $U/t=160$, obtained from PQMC, and (c) is exactly at the $U=\infty$ limit, obtained from IQMC. (d) The dynamical spin spectrum of free $\pi$-flux Dirac fermions. The lower and upper dashed curves in (b), (c) and (d) trace out the lower and upper edges of the two-spinon continuum, i.e. $\omega^{\text{lower}}(\mathbf{q}) \propto \min_{\mathbf{k}\in \text{BZ}} | \epsilon(\mathbf{k}) + \epsilon(\mathbf{q-k})|$ and $\omega^{\text{upper}}(\mathbf{q}) \propto \max_{\mathbf{k}\in \text{BZ}} | \epsilon(\mathbf{k}) + \epsilon(\mathbf{q-k})|$, assuming free spinons with the $\pi$-flux state dispersion. The calculation of the free two-spinon continuum is shown in Sec.~\ref{app:B} in Appendix.}
	\label{fig:spectrum}
\end{figure*}

\section{Physical observables for the Dirac QSL}
The confirmation of emergent Dirac spinon coupled with U(1) gauge field with power-law correlation functions for bond and spin operators, are still too abstract from the experimental point of view. To this end, we follow the tradition in condensed matter experiments to further probe the response of the system by external parameters. Here we focus on the thermodynamic and dynamic measurements at the Dirac QSL phase.
It was proposed that from the temperature dependence of the magnetic susceptibility, one could expect a $\chi_0\sim T$ behavior of such U(1) Dirac spin liquid state at low temperatures~\cite{kimMassless1997,ranProjected2007}, distinctively different from the $\chi_0 \sim T^{-1}$ for a Pauli paramagnet with Fermi surface. We define dynamic spin correlation function as
$\chi_S(\mathbf{q},\tau) \equiv \frac{1}{L^2}\sum_{\mathbf{r}} e^{-i\mathbf{q}\cdot\mathbf{r}}\sum_{\alpha,\beta}\langle S^{\alpha}_{\beta}(i,\tau) S^{\beta}_{\alpha}(j,0)\rangle$ with $\tau\in[0,\beta]$, then the susceptibility can be computed as $\chi_0(T) = \int^{\beta=\frac{1}{T}}_{0} \chi_{S}(\mathbf{q}=0,\tau)d\tau$.
As shown in Fig.~\ref{fig:largeU}~(b), $\chi_0(T)$, obtained from finite temperature QMC, obviously violate $\chi_0\sim T$ behavior at $U/t=56$, where the cVBS order is very strong and the state is gapped at low temperature.
However, as increasing plaquette interaction weakens the cVBS order, at $U/t=120$ and $U/t=160$, $\chi_0\sim T$ behavior gradually emerges.
And at $U=\infty$, i.e. in the Dirac QSL, $\chi_0(T)$ decays as a function of temperature and is in good agreement with linear behavior.

Moreover, with the good quality dynamic spin correlation function data $\chi_S(\mathbf{q},\tau)$ at hand, the stochastic analytic continuation (SAC) scheme can reveal reliable spectral information, $S(\mathbf{q},\omega)$, as have been widely tested in fermionic and bosonic quantum many-body systems in 1D, 2D and 3D~\cite{sandvikStochastic1998,beachIdentifying2004,syljuasenUsing2008,shaoNearly2017,sandvikConstrained2016,sunDynamical2018,wangDynamics2019,wangVestigial2021,wangFractionalized2021,yanTopological2021,zhouAmplitude2021,jiangPseudogap2021,yanTriangular2022,shaoProgress2022,zhouEvolution2022,panDynamical2022,maDynamical2018}, and compared with both neutron scattering and NMR experiments~\cite{liKosterlitz2020,huEvidence2020} and the exact solution and exact diagonalization numerics~\cite{shaoNearly2017,zhouAmplitude2021,panDynamical2022,zhangSuperconductivity2021}. We therefore compute the spin spectrum $S(\mathbf{q},\omega)$ with QMC+SAC at different $U$, the spectra are shown in Fig.~\ref{fig:spectrum}.

By comparison with spin spectra at $U/t=56$ [Fig.~\ref{fig:spectrum} (a)], $U/t=160$ [Fig.~\ref{fig:spectrum} (b)], $U=\infty$ [Fig.~\ref{fig:spectrum} (c)] and analytical calculation of free $\pi$-flux model on square lattice [Fig.~\ref{fig:spectrum} (d)], several exotic features of the spectra can
be unambiguously identified. First, we observe broad and
prominent continua in $S(\mathbf{q},\omega)$ at the $U/t=160$ and $U=\infty$, which reflect the expected fractionalization and emergence of deconfined spinons. In contrast, inside the cVBS phase, the $S(\mathbf{q},\omega)$ is gapped due to the translational symmetry breaking.

Second, we find that the lower edges
of both $U/t=160$ and $U=\infty$ spectra can be well accounted for by a remarkably simple dispersion relation,
$\omega^{\text{lower}}(\mathbf{q}) \propto \min_{\mathbf{k}\in \text{BZ}} | \epsilon(\mathbf{k}) + \epsilon(\mathbf{q-k})|$
[the lower dashed lines in Fig.~\ref{fig:spectrum} (b) and (c)], which matches the dispersion relation of free fermionic spinon in the square lattice
$\pi$-flux state [the lower dashed lines in Fig.~\ref{fig:spectrum} (d)]~\cite{maDynamical2018,wangDynamics2019}. This points us to the cQED$_3$ description of the Dirac QSL state above~\cite{xuMonte2019}, which is also proposed as the low-energy description of the deconfined quantum critical point~\cite{senthilDeconfined2004,senthilQuantum2004}. If indeed the broad spectral
functions seen in Fig.~\ref{fig:spectrum} (b) and (c) are due to two independently propagated spinons, 
the upper spectral bound can also be obtained by
$\omega^{\text{upper}}(\mathbf{q}) \propto \max_{\mathbf{k}\in \text{BZ}} | \epsilon(\mathbf{k}) + \epsilon(\mathbf{q-k})|$
[the upper dashed lines in Fig.~\ref{fig:spectrum} (b), (c) and (d)].
This is only in reasonable agreement with the observed distribution of the main spectral
weight, though clearly some weights arising from spinon interactions are also present at higher energies. 

Furthermore, gapless continua observed at both $\mathbf{X}_1$ and $\mathbf{M}$ points are consistent with the fact in Fig.~\ref{fig:largeU} (c) and (d) both AF (staggered) spin and columnar VBS correlations exhibiting the same power-law decay. The same observation of gapless spin continua at $(\pi,\pi)$ and $(\pi,0)$ have also been confirmed in the $SU(2)$ J-Q model at its deconfined quantum critical point~\cite{maDynamical2018}. It was also argued recently that these spectral continua in magnetic models and materials, are evidence for deconfined quantum critical point, in metal-organic compound Cu(DCOO)$_2 \cdot 4$D$_2$O and in the Shastry-Sutherland compound SrCu$_2$(BO$_3$)$_2$~\cite{shaoNearly2017,dallaFractional2015,guoQuantum2020,guangyuEmergent2021,cuiDeconfined2022}. Such spectral signatures can certainly be probed in neutron scattering, RIXS, NMR and scanning tunnelling spectroscopy techniques.

\section{Discussion}
As mentioned throughout the paper, the coupling between fermionic matter and gauge fields is of fundamental importance in both high-energy and condensed-matter physics. In the latter, gauge fields can emerge as a consequence of fractionalization in quantum materials, which may be realizable in certain frustrated magnets, such as the recent triangular antiferromagnets ${\mathrm{Na Yb O}}_{2}$~\cite{dingGapless2019} and $\mathrm{Sr}_3\mathrm{CuSb}_2\mathrm{O}_9$~\cite{Kundu2020}, and kagome antiferromagnet ${\mathrm{YCu}}_{3}{(\mathrm{OH})}_{6}{\mathrm{Br}}_{2}[{\mathrm{Br}}_{x}{(\mathrm{OH})}_{1\ensuremath{-}x}]$~\cite{zengPossible2022}. At the model level, prominent proposals have been applied to high-temperatur superconductors~\cite{wenTheroy1996,kimMassless1997,leeDoping2006}, QSL~\cite{marstonInstantons1990,leeU12005,motrunichVariational2005,ranProjected2007} and deconfined quantum critical points~\cite{senthilQuantum2004,sandvikEvidence2007,nahumDeconfined2015,qinDuality2017,maDynamical2018,maRole2019}. But since these strongly correlated quantum states are characterized by topological order or coupled matter fields and gauge fields, it is often-times difficult to unambiguously identify them in simulation with the relevant low-energy fractionalized excitations and their characteristic properties.

In this work, we have overcome these difficulties with model design and numerical methodology development. By means of three different and complementary QMC simulation techniques, we reveal the phase diagram of correlated $SU(4)$ Dirac fermions on the $\pi$-flux square lattice with plaquette interaction. We find a Gross-Neveu QCP with emergent U(1) symmetry separating the massless Dirac fermions and an columnar VBS at finite interaction, and a possible Dirac QSL at the infinite interaction limit with its characteristic thermodynamic and dynamic properties accessible to experiments. Such unexpected sequence of novel quantum states in the simple-looking model, unify the key ingredients including emergent symmetry~\cite{xuKekule2018,liaoValence2019,janssenConfinement2020,liaoGross2022}, deconfined fractionalization~\cite{xuMonte2019,maDynamical2018,maRole2019} and the dynamic coupling between emergent matter and gauge fields~\cite{heQuantum2016,xuMonte2019,wangDynamics2019}. Our work therefore provides a solid foundation for the future exploration of the novel quantum matter originated from the interplay of the low-energy relativistic dispersion and strong extended and long-range interactions.

\section*{Acknowledgements }
We acknowledge Zheng Yan and Chong Wang for valuable discussions on the subject. Y.D.L. acknowledges the support of Project funded by China Postdoctoral Science Foundation through Grants No. 2021M700857 and No. 2021TQ0076.
X.Y.X. is sponsored by the National Key R\&D Program of China (Grant No. 2021YFA1401400), Shanghai Pujiang Program under Grant No. 21PJ1407200, Yangyang Development Fund, and startup funds from SJTU. \
Z.Y.M. acknowledges support from the Research Grants Council of Hong Kong SAR of China (Grant Nos. 17303019, 17301420, 17301721, AoE/P-701/20 and 17309822), the GD-NSF (No.2022A1515011007), the K. C. Wong Education Foundation (Grant No. GJTD-2020-01) and the Seed Funding “Quantum-Inspired explainable-AI” at the HKU-TCL Joint Research Centre for Artificial Intelligence.
Y.Q. acknowledges support from the the National Natural Science Foundation of China (Grant Nos. 11874115 and 12174068).
The authors also acknowledge \href{https://www.paratera.com/}{Beijng PARATERA Tech Co.,Ltd.} for providing HPC resources that have contributed to the research results reported within this paper.

\appendix

\section{Methods}
\label{app:A}
\subsection{Finite temperature auxiliary field QMC method}
Here, we represent Hamiltonian as $H=H_0+H_U$ with non-interacting part $H_0=-\sum_{\langle i j\rangle, \alpha} t_{i j}\left(c_{i \alpha}^{\dagger} c_{j \alpha}+\text { H.c. }\right)$ and interacting part $H_U=U \sum_{\square}\left(n_{\square}-2\right)^{2}$. 
Since $H_0$ and $H_U$ do not commute, we use Trotter decomposition to separate $H_0$ and $H_U$ in the imaginary time propagation
\begin{equation}
	Z=\operatorname{Tr}\left[\mathrm{e}^{-\beta H}\right]=\operatorname{Tr}\left[\left(\mathrm{e}^{-\Delta_{\tau} H_{U}} \mathrm{e}^{-\Delta_{\tau} H_{0}}\right)^{M}\right]+\mathcal{O}\left(\Delta_{\tau}^{2}\right),
\end{equation}
where $Z$ is the partition function, $\beta=M\Delta\tau$ is the inverse temperature.
In QMC method, we can only deal with quadratic fermionic operator, while $H_U$ contains the quartic term, thus we should employ a Hubbard-Stratonovich (HS) decomposition as
\begin{equation}
e^{-\Delta\tau U(n_{\square}-2)^{2}}=\frac{1}{4}\sum_{\{s_{\square,\tau}\}}\gamma(s_{\square,\tau})e^{-2\alpha\eta(s_{\square,\tau})}e^{\alpha\eta(s_{\square,\tau})n_{\square}}
\label{eq:decompo}
\end{equation}
with $\alpha=\sqrt{-\Delta\tau U}$, $\gamma(\pm1)=1+\sqrt{6}/3$,
$\gamma(\pm2)=1-\sqrt{6}/3$, $\eta(\pm1)=\pm\sqrt{2(3-\sqrt{6})}$,
$\eta(\pm2)=\pm\sqrt{2(3+\sqrt{6})}$ and the sum symbol is taken over the auxiliary fields $s_{\square,\tau}$ on each $\tau$-th time slice square plaquette. Now, the interacting part is transformed into quadratic term but coupled with an auxiliary field.
Following simulations are based on the single-particle basis $\boldsymbol{c} = \{c_1, c_2 \cdots c_{N} \}$, so we can use the matrix notation $K$ and $V$ to represent $H_0$ and $H_U$ operators. 
We define the imaginary time propagators 
\begin{equation}
\begin{aligned}
&U_{s_{\square}}\left(\tau_{2}, \tau_{1}\right)=\prod_{m=m_{1}+1}^{m_{2}} e^{\boldsymbol{c}^{\dagger} V\left(s_{\square,m\Delta\tau}\right) \boldsymbol{c}} e^{-\Delta_{\tau} \boldsymbol{c}^{\dagger} K \boldsymbol{c}}, \\
&B_{s_{\square}}\left(\tau_{2}, \tau_{1}\right)=\prod_{m=m_{1}+1}^{m_{2}} e^{V\left(s_{\square,m\Delta\tau}\right)} e^{-\Delta_{\tau} K},
\end{aligned}
\end{equation}
where $m_1 \Delta\tau=\tau_1$ and $m_2 \Delta\tau=\tau_2$. 
Then partition funciton $Z$ can be rewritten as 
\begin{widetext}
\begin{equation}
Z=\sum_{\{s_{\square,\tau} \}} \operatorname{Tr}\left[U_{s_{\square}}(\beta, 0)\right] \prod_{m=1}^{M} \gamma(s_{\square,m\Delta\tau})e^{-2\alpha\eta(s_{\square,m\Delta\tau})} = \sum_{\{s_{\square,\tau} \}} \operatorname{det}\left[1+B_{s_{\square}}(\beta, 0)\right] \prod_{m=1}^{M} \gamma(s_{\square,m\Delta\tau})e^{-2\alpha\eta(s_{\square,m\Delta\tau})}.
\end{equation}
\end{widetext}
Physical observables are measured according to 
\begin{equation}
\langle O \rangle = \frac{\operatorname{Tr}\left[\mathrm{e}^{-\beta H} O\right]}{\operatorname{Tr}\left[\mathrm{e}^{-\beta H}\right]}.
\end{equation}
The equal-time single-particle Green function $G_{i,j}(\tau,\tau)$ is given by
\begin{equation}
\left\langle c_{i,\tau} c_{j,\tau}^{\dagger}\right\rangle=\left(1+B_{s_{\square}}(\tau, 0) B_{s_{\square}}(\beta, \tau)\right)_{i, j}^{-1},
\end{equation}
and the dynamical single-particle Green function $G_{i,j}(\tau_1,\tau_2)$ is given by
\begin{equation}
\left\langle c_{i,\tau_1} c_{j,\tau_2}^{\dagger}\right\rangle=-\left[\left(\mathbf{1}-\mathbf{G}\left(\tau_{1}, \tau_{1}\right)\right) B_{s_{\square}}^{-1}\left(\tau_{2}, \tau_{1}\right)\right]_{i, j}.
\end{equation}
Other physical observables can be calculated from single-particle Green function through Wick's theorem. More technical details of the finite-temperature QMC algorithms can be found in the reference book~\cite{assaadWorld-line2008}.

\subsection{Projection QMC method}
Since we also want to investigate the ground state properties, the projection quantum Monte Carlo (PQMC) method is a good choice. 
We can calculate the ground state wave function $\vert \Psi_0 \rangle$ through the projection of a trial wave function $\vert \Psi_T \rangle$ as $\vert \Psi_0 \rangle = \lim\limits_{\Theta \to \infty} e^{-\frac{\Theta}{2} \mathbf{H}} \vert \Psi_T \rangle$, where $\Theta$ is the projection length. 
Actually, in PQMC method, we use $\langle\Psi_{T}|e^{-\Theta H}|\Psi_{T}\rangle$ to replace the role of partition function in finite-temperature version.
Physical observables $O$ can be measured according to
\begin{widetext}
\begin{equation}
\label{eq:observablepqmc}
\langle O \rangle = \frac{\langle \Psi_0 \vert O \vert \Psi_0 \rangle}{\langle \Psi_0 \vert \Psi_0 \rangle} 
						= \lim\limits_{\Theta \to \infty} \frac{\langle \Psi_T \vert  e^{-\frac{\Theta}{2} \mathbf{H}} O  e^{-\frac{\Theta}{2} \mathbf{H}} \vert \Psi_T \rangle}{\langle \Psi_T \vert  e^{-\Theta \mathbf{H}} \vert \Psi_T \rangle} .
\end{equation}
After the Trotter and HS decomposition, we have
\begin{equation}
\langle\Psi_{T}|e^{-\Theta H}|\Psi_{T}\rangle=\sum_{\{s_{\square,\tau}\}}\left[\left(\prod_{\tau}\prod_{\square}\gamma(s_{\square,\tau})e^{-2\alpha\eta(s_{\square,\tau})}\right)\det\left[P^{\dagger}B(\Theta,0)P\right]\right]
\label{eq:mcweight}
\end{equation}
\end{widetext}
here $P$ is the coefficient matrix of trial wave function $|\Psi_T\rangle$; $B$ matrix is defined as
$B(\tau+1,\tau)=e^{V}e^{-\Delta_\tau K}$
and has a property $B(\tau_3,\tau_1)=B(\tau_3,\tau_2)B(\tau_2,\tau_1)$.
In the practice,  we choose the ground state wavefunction of the half-filled non-interacting parts of Hamiltonian as the trial wave function. 
The Monte Carlo sampling of auxiliary fields are further performed based on the weight defined in the sum of  Eq.~\eqref{eq:mcweight}. Single particle observables are measured by Green's function directly and many body correlation functions are measured from the products of single-particle Green's function based on their corresponding form after Wick-decomposition. 
The equal time Green's function are calculated as
\begin{equation}
G(\tau,\tau)=1-R(\tau)\left(L(\tau)R(\tau)\right)^{-1}L(\tau)
\end{equation}
with $R(\tau)=B(\tau,0)P$, $L(\tau)=P^{\dagger}B(\Theta,\tau)$. 
The imaginary-time displaced Green's function  
$G(\tau,0) \equiv \left\langle \mathbf{c} \left(\frac{\tau}{2}\right) \mathbf{c}^{\dagger}\left(-\frac{\tau}{2}\right)\right\rangle$
are calculated as
\begin{equation}
\left\langle \mathbf{c} \left(\frac{\tau}{2}\right) \boldsymbol{c}^{\dagger}\left(-\frac{\tau}{2}\right)\right\rangle=\frac{\left\langle\Psi_{T}\left|e^{-\left(\frac{\Theta}{2}+\frac{\tau}{2}\right) H} \boldsymbol{c} e^{-\tau H} \boldsymbol{c}^{\dagger} e^{-\left(\frac{\Theta}{2}-\frac{\tau}{2}\right) H}\right| \Psi_{T}\right\rangle}{\left\langle\Psi_{T}\left|e^{-\Theta H}\right| \Psi_{T}\right\rangle}
\end{equation}
More technique details of PQMC method, please also refer to Refs~\cite{assaadWorld-line2008}. 

\subsection{Infinite $U$ QMC algorithm}
For infinite $U$ calculation, we use following formula to perform the $SU(N)$ infinite $U$ projection~\cite{ouyangProjection2021}
\begin{equation}
 \left.e^{-\Delta_\tau U (n_{\square}- 2)^2}\right|_{U\rightarrow+\infty} 
=  \frac{1}{M}\sum_{s_{\square,\tau}=1}^M e^{\frac{\text{i} 8\pi s_{\square,\tau}}{M} (n_{\square}- \frac{N}{2})},
\end{equation}
where we can set $M=2N+1$.
As the infinite $U$ term has now been replaced by fermion bilinears coupled to auxiliary fields, we can further use above finite temperature and projection QMC scheme to perform the calculation.

\subsection{SAC method }
\begin{figure}[htb!]
\includegraphics[width=1\columnwidth]{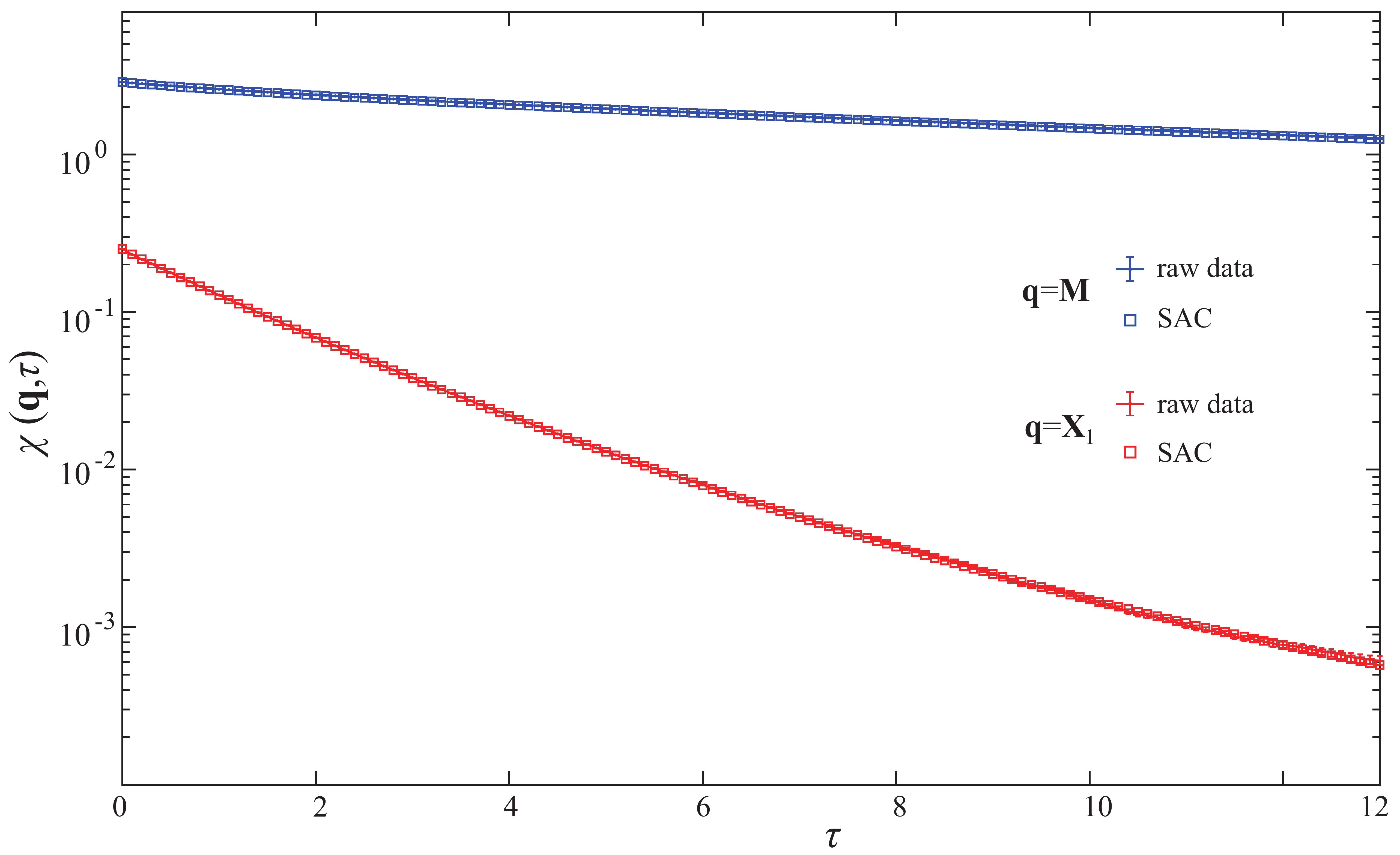}
\caption{The log imaginary-time dynamic spin correlations at $\mathbf{M}$ and $\mathbf{X}_1$ momentum points as function of $\tau$, obtained from PQMC method and the Laplace transformation of the spin spectrum $S(\mathbf{q},\omega)$, respectively. Here $U=\infty$ and $L=20$. }
\label{fig:SAC}
\end{figure}

The SAC method could help us obtain the spin spectrum $S(\mathbf{q},\omega)$ from imaginary-time dynamic spin correlation $\chi_S(\mathbf{q},\tau)$. 
There is a regular relation between $S(\mathbf{q},\omega)$ and $\chi_S(\mathbf{q},\tau)$
\begin{equation}\label{eq:sac}
S(\mathbf{q},\tau)=\int_{0}^{\infty} d \omega \chi_S(\mathbf{q},\omega) K(\tau, \omega),
\end{equation}
here $K(\tau, \omega)=\frac{1}{\pi}\left(e^{-\tau \omega}+e^{-(\beta-\tau) \omega}\right)$ is known as Kernel function for Bosonic particle at finite-temperature. 
At zero-temperature, we could just set $\beta\rightarrow \infty$ simply. 
The more details of SAC method could be found in previous studies~\cite{sandvikStochastic1998,beachIdentifying2004,syljuasenUsing2008,shaoNearly2017,sandvikConstrained2016,
sunDynamical2018,wangDynamics2019,wangVestigial2021,wangFractionalized2021,yanTopological2021,zhouAmplitude2021,
jiangPseudogap2021,yanTriangular2022,shaoProgress2022,zhouEvolution2022,panDynamical2022,maDynamical2018}, we won't repeat at here.
While, as shown in Fig.~\ref{fig:SAC}, we plot the raw imaginary-time dynamic spin correlation $\chi_S(\mathbf{q},\tau)$ at $\mathbf{q}=\mathbf{M}$ and $\mathbf{X}_1$ with the same quantity obtained from the Laplace transformation of the spin spectrum $S(\mathbf{q},\omega)$ according to Eq.~\eqref{eq:sac}, and the comparison is perfect.

\section{Other QMC data}
\label{app:B}
\subsection{The single-particle gap and VBS order parameter near QCP of SM-VBS }
\begin{figure*}[htb!]
\includegraphics[width=1.65\columnwidth]{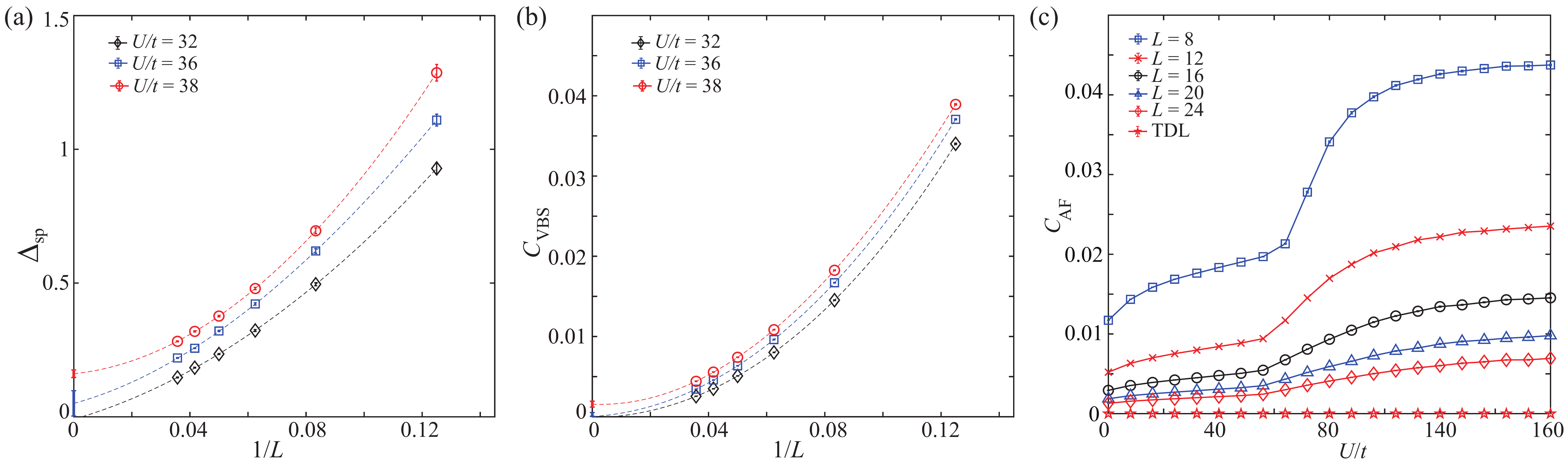}
\caption{(a) $1/L$ extrapolation of single-particle gap. (b) $1/L$ extrapolation of VBS structure factor. The dash lines are obtained by cubic polynomial fitting with least squares method. (c) The AF structure factors as function of interaction strength. Error bars are smaller than the symbols.
These data are obtained from PQMC method.}
\label{fig:AF-order}
\end{figure*}

The Dirac SM to cVBS phase transition is a transition from massless Dirac fermion to insulator, thus single-particle gap $\Delta_{\text{sp}}$ will open at Dirac cones, i.e. at the momentum point $\mathbf{K}_0$.
We can extract $\Delta_{\mathrm{sp}}(\mathbf{K}_0,L)$ from a fit to the asymptotic long imaginary time behavior of the single-particle Green's function $G(\mathbf{k}, \tau, L) \propto e^{-\Delta_{\mathrm{sp}}(\mathbf{k},L) \tau}$, here
\begin{equation}
G(\mathbf{k}, \tau, L)=\frac{1}{L^{4}} \sum_{i, j, \sigma} e^{i \mathbf{k} \cdot\left(\mathbf{r}_{i}-\mathbf{r}_{j}\right)}\left\langle c_{i, \sigma}^{\dagger}(\tau) c_{j, \sigma}(0)\right\rangle.
\end{equation}
Then extrapolate it to the thermodynamic limit (TDL).
As shown in Fig.~\ref{fig:AF-order}~(a), it is clear that $\Delta_{\mathrm{sp}}(\mathbf{K}_0,L\rightarrow\infty) \rightarrow 0$ at $U/t=32$ in the SM phase, $\Delta_{\mathrm{sp}}(\mathbf{K}_0,L\rightarrow\infty) > 0$ at $U/t=38$ in the VBS phase, and  $\Delta_{\mathrm{sp}}(\mathbf{K}_0,L\rightarrow\infty) \sim 0$ at $U_c/t=36$.
The $1/L$ extrapolation of $C_{\text{VBS}}$ has the similar behavior as $\Delta_{\mathrm{sp}}(\mathbf{K}_0,L)$ near $U_c$, as shown in Fig.~\ref{fig:AF-order}~(b).

\subsection{The absence of AF order}
In Hubbard-like model, the AF order usually dominate when VBS order become weak in strong coupling region~\cite{xuKekule2018,liaoGross2022}. 
Here, we demonstrate that the AF order is absent in our model. 
We define the structure factor of AF order as 
$
C_\text{AF} (\mathbf{k}, L) \equiv \frac{1}{L^4}\sum_{\mathbf{r}} e^{-i\mathbf{q}\cdot\mathbf{r}}\sum_{\alpha,\beta}\langle S^{\alpha}_{\beta}(i) S^{\beta}_{\alpha}(j)\rangle
$, $C_\text{AF}$ will peak at momentum point $\mathbf{M}$ for AF order.
We can perform $1/L$ extrapolation of $C_\text{AF}(\mathbf{M},L)$ to get the AF structure factor at TDL, markded as $C_\text{AF}(\mathbf{M},\infty)$.
As shown in Fig.~\ref{fig:AF-order} (c), $C_\text{AF}(\mathbf{M},\infty)$ vanish at whole interaction strength range $U/t \in [0,160]$, which mean the AF order is absent in our model.

\begin{figure}[htb!]
\includegraphics[width=1\columnwidth]{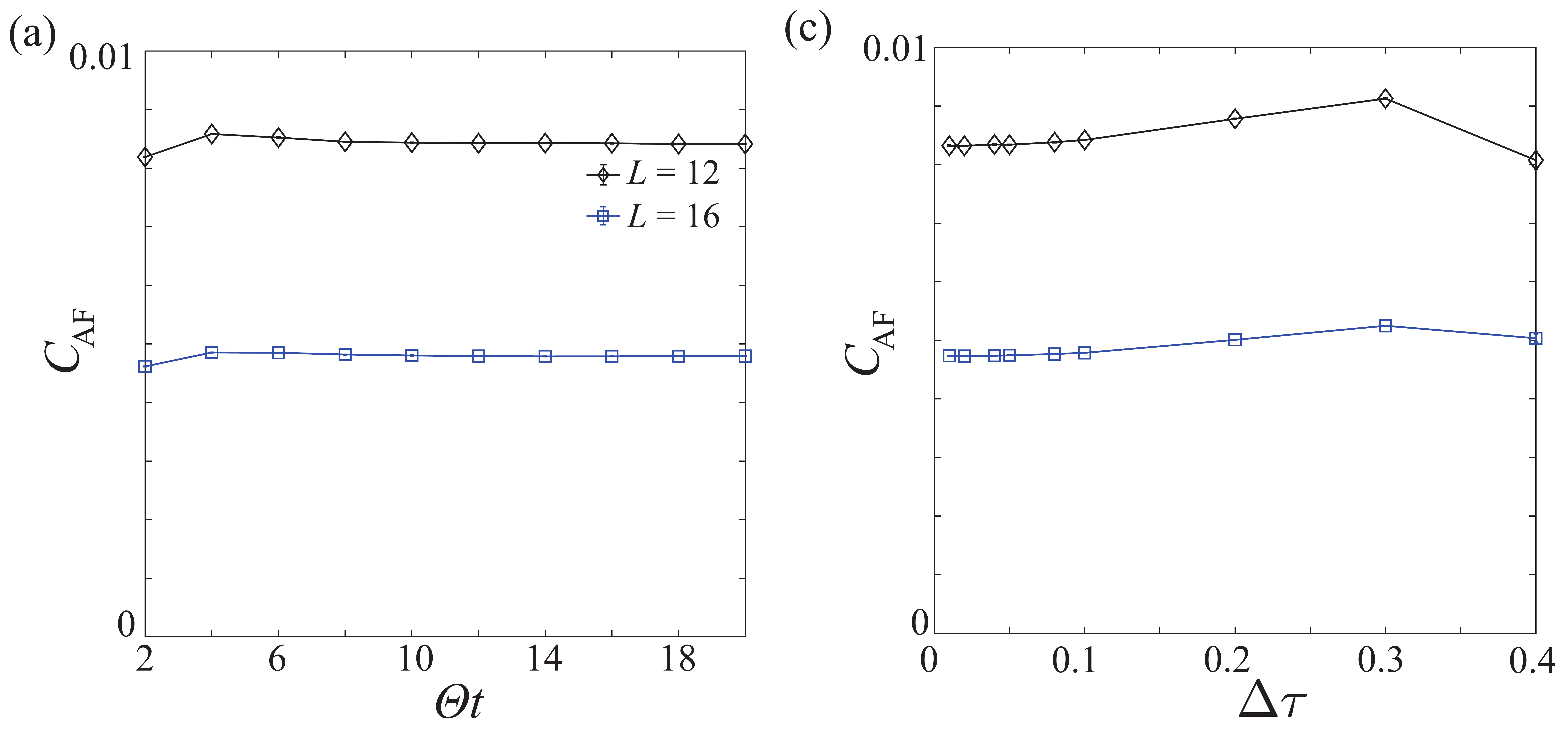}
\caption{The structure factors of AF order with respect to (a) with respect to the projection length $\Theta$ with $\Delta \tau=0.1$ and (b) the time slice $\Delta\tau$ with $\Theta t = L$. These results are obtained at $U/t = 40$ near $U_c$ with PQMC method. All lines in the figure are guides to the eye.}
\label{fig:tau}
\end{figure}

As mentioned in our paper, we set projection time $\Theta t = L$ for equal-time measurement, $\Theta t = L+20$ for imaginary-time measurement and discrete the time slice $\Delta\tau=0.1$ in PQMC method, as shown in Fig.~\ref{fig:tau}, we have confirmed that this set-up is enough to achieve convergent and error controllable $C_{AF}$ for our model.

\section{Dynamical spin spectrum of free $\pi$-flux model}
We have a free $SU(4)$ $\pi$-flux model on square lattice with Hamiltonian
\begin{equation}
	H_0=- \sum_{\langle i j\rangle, \alpha} t_{i j}\left(c_{i \alpha}^{\dagger} c_{j \alpha}+\text { H.c. }\right).
\end{equation}
, and consider two-site unit cell, with inner cell coordinates $\mathbf{v}_{1}=(0,0)$,
$\mathbf{v}_{2}=(1,0)$.
We transform the Hamiltonian to the momentum space, 
\begin{equation}
H_0=-t\sum_{\mathbf{k}}c_{\mathbf{k}}^{\dagger}\left(\begin{array}{cc}
2\cos k_y & 1+e^{-2ik_{x}}\\
1+e^{2ik_{x}} & -2\cos k_y
\end{array}\right)c_{\mathbf{k}}
\end{equation}
where $c_{\mathbf{k}}=(c_{1,\mathbf{k}},c_{2,\mathbf{k}})^{T}$. We further write
$H$ in the diagonal basis
\begin{align}
H_0 & =\sum_{\mathbf{k},a,b,m,n}c_{a,\mathbf{k}}^{\dagger}(U_{\mathbf{k}})_{am}(D_{\mathbf{k}})_{mn}(U_{\mathbf{k}}^{-1})_{nb}c_{b,\mathbf{k}}\nonumber \\
 & \equiv\sum_{\mathbf{k},a}\epsilon_{a,\mathbf{k}}f_{a,\mathbf{k}}^{\dagger}f_{a,\mathbf{k}}
\end{align}
where we use $a,b,m,n\in [1,2] $ to denote sublattice index, $(D_{\mathbf{k}})_{mn}=\delta_{mn}\epsilon_{m,\mathbf{k}}$, $\epsilon_{m,\mathbf{k}}=\pm 2 \sqrt{\cos^2 k_x + \cos^2 k_y}$ is the dispersion,
and we define $f_{a,\mathbf{k}}\equiv U_{\mathbf{k}}^{-1}c_{a,\mathbf{k}}$. Here the dispersion
is flavor degenerate, and we have omitted the flavor index, . The sum
is over $\mathbf{k}$ in the small (rectangle) BZ, i.e. $k_{x}\in(-\frac{\pi}{2},\frac{\pi}{2})$ and $k_{y}\in(-\pi,\pi$).

To calculate spin spectrum, we can use fluctuation-dissipation theorem,
which relates the spin spectrum to the spin susceptibility. The spin
susceptibility has following form

\begin{equation}
\chi_{ab}^{R}(\mathbf{r}-\mathbf{r}',t-t')=-i\theta(t-t')\sum_{\mu\nu}\langle[S_{\nu}^{\mu}(\mathbf{r},a,t),S_{\mu}^{\nu}(\mathbf{r}',b,t')]\rangle .
\end{equation}
Fourier transform
to momentum space, we get
\begin{equation}
\chi_{ab}^{R}(\mathbf{q},t-t')=-i\theta(t-t')\frac{1}{V}\sum_{\mu\nu}\langle[S_{\nu}^{\mu}(\mathbf{q},a,t),S_{\mu}^{\nu}(-\mathbf{q},b,t')]\rangle
\end{equation}
, where $S^{\alpha}_{\beta}(\mathbf{k},a,t)=c^{\dagger}_{\mathbf{k},\alpha}c_{\mathbf{k},\beta}-\frac{\delta_{\alpha,\beta}}{4}\sum_{\gamma}c^{\dagger}_{\mathbf{k},\gamma}c_{\mathbf{k},\gamma}$ are the $SU(4)$ spin full operators with $\alpha,\beta,\gamma \in [1,4]$.

For non-interacting $SU(4)$ case, the time dependence of spin operator is
given by
\begin{widetext}
\begin{align}
S_{\nu}^{\mu}(\mathbf{q},a,t) & =\sum_{\mathbf{k}}e^{iHt}(c_{a,\mathbf{k},\mu}^{\dagger}c_{a,\mathbf{k}+\mathbf{q},\nu}-\frac{\delta_{\mu\nu}}{N}\sum_{\lambda}c_{a,\mathbf{k},\lambda}^{\dagger}c_{a,\mathbf{k}+\mathbf{q},\lambda})e^{-iH_0t}\nonumber \\
 & =\sum_{\mathbf{k}}\sum_{mn}(U_{\mathbf{k}}^{-1})_{ma}(U_{\mathbf{k}+\mathbf{q}})_{an}(f_{m,\mathbf{k},\mu}^{\dagger}f_{n,\mathbf{k}+\mathbf{q},\nu}-\frac{\delta_{\mu\nu}}{N}\sum_{\rho}f_{m,\mathbf{k},\rho}^{\dagger}f_{n,\mathbf{k}+\mathbf{q},\rho})e^{i(\epsilon_{m,\mathbf{k}}-\epsilon_{n,\mathbf{k}+\mathbf{q}})t}
\end{align}
Therefore
\begin{align}
\chi_{ab}^{R}(\mathbf{q},t-t')= & -i\theta(t-t')\frac{N^{2}-1}{V}\sum_{\mathbf{k}}\sum_{mn}(U_{\mathbf{k}}^{-1})_{ma}(U_{\mathbf{k}+\mathbf{q}})_{an}(U_{\mathbf{k}+\mathbf{q}}^{-1})_{nb}(U_{\mathbf{k}})_{bm} [n_{F}(\epsilon_{m,\mathbf{k}})-n_{F}(\epsilon_{n,\mathbf{k}+\mathbf{q}})]e^{i(\epsilon_{m,\mathbf{k}}-\epsilon_{n,\mathbf{k}+\mathbf{q}})(t-t')}
\end{align}
Perform Fourier transformation to frequency space, we get
\begin{align}
\chi_{ab}^{R}(\mathbf{q},\omega)
= \frac{N^2-1}{V}\sum_{\mathbf{k}}\sum_{mn}(U_{\mathbf{k}}^{-1})_{ma}(U_{\mathbf{k}+\mathbf{q}})_{an}(U_{\mathbf{k}+\mathbf{q}}^{-1})_{nb}(U_{\mathbf{k}})_{bm}\frac{n_{F}(\epsilon_{m,\mathbf{k}})-n_{F}(\epsilon_{n,\mathbf{k}+\mathbf{q}})}{\epsilon_{m,\mathbf{k}}-\epsilon_{n,\mathbf{k}+\mathbf{q}}+\omega+i\eta}
\end{align}

To get the full spin spectrum, we should keep in mind that above derivation
is in small BZ. Now we consider the spin correlation on a square lattice
(one site unit cell), and relate it to above formula.

\begin{equation}
\chi^{R}(\mathbf{r}_{i}-\mathbf{r}_{j},t-t')=-i\theta(t-t')\sum_{\mu\nu}\langle[S_{\nu}^{\mu}(\mathbf{r}_{i},t),S_{\mu}^{\nu}(\mathbf{r}_{j},t')]\rangle
\end{equation}
denote $\mathbf{r}_{i}=\mathbf{r}+\mathbf{v}_{a}$, $\mathbf{r}_{j}=\mathbf{r}'+\mathbf{v}_{b}$, we have
\begin{equation}
\chi^{R}(\mathbf{r}+\mathbf{v}_{a}-\mathbf{r}'-\mathbf{v}_{b},t-t')=-i\theta(t-t')\sum_{\mu\nu}\langle[S_{\nu}^{\mu}(\mathbf{r},a,t),S_{\mu}^{\nu}(\mathbf{r}',b,t')]\rangle\equiv\chi_{ab}^{R}(\mathbf{r}-\mathbf{r}',t-t')
\end{equation}
\end{widetext}
Perform fourier transformation, we get following relation
\begin{equation}
\chi^{R}(\mathbf{q},\omega)=\sum_{ab}\chi_{ab}^{R}(\mathbf{q},\omega)e^{-i\mathbf{q}(\mathbf{v}_{a}-\mathbf{v}_{b})},
\end{equation}
and the full spin spectrum is
\begin{equation}
S(\mathbf{q},\omega)=-2\text{Im}\chi^{R}(\mathbf{q},\omega)
\end{equation}

\bibliographystyle{apsrev4-2}
\bibliography{main}

\end{document}